\documentclass[manuscript=article]{achemso}

\setkeys{acs}{usetitle=true}
\usepackage{subfigure}
\usepackage{amsmath}
\usepackage[numbers]{natbib}
\usepackage[version=3]{mhchem}
\usepackage{multirow}
\usepackage{xr}
\usepackage{xcolor}
\usepackage{longtable}
\usepackage{array}
\usepackage{color, colortbl}
\definecolor{Gray}{gray}{0.9}
\usepackage{float}
\usepackage{bm}
\usepackage{longtable}
\usepackage{array}

\newcolumntype{P}[1]{>{\centering\arraybackslash}p{#1}}
\newcolumntype{M}[1]{>{\centering\arraybackslash}m{#1}}

\title{Complex Strengthening Mechanisms in the NbMoTaW Multi-Principal Element Alloy}
\author{Xiang-Guo Li}
\affiliation[UCSD]{Department of NanoEngineering, University of California San Diego, 9500 Gilman Dr, Mail Code 0448, La Jolla, CA 92093-0448, United States}
\author{Chi Chen}
\affiliation[UCSD]{Department of NanoEngineering, University of California San Diego, 9500 Gilman Dr, Mail Code 0448, La Jolla, CA 92093-0448, United States}
\author{Hui Zheng}
\affiliation[UCSD]{Department of NanoEngineering, University of California San Diego, 9500 Gilman Dr, Mail Code 0448, La Jolla, CA 92093-0448, United States}
\author{Yunxing Zuo}
\affiliation[UCSD]{Department of NanoEngineering, University of California San Diego, 9500 Gilman Dr, Mail Code 0448, La Jolla, CA 92093-0448, United States}
\author{Shyue Ping Ong}
\email{ongsp@eng.ucsd.edu}
\affiliation[UCSD]{Department of NanoEngineering, University of California San Diego, 9500 Gilman Dr, Mail Code 0448, La Jolla, CA 92093-0448, United States}
\date{}

\begin{document}

\maketitle

\begin{abstract}
Refractory multi-principal element alloys (MPEAs) have exceptional mechanical properties, including high strength-to-weight ratio and fracture toughness, at high temperatures. Here, we elucidate the complex interplay between segregation, short range order and strengthening in the NbMoTaW MPEA through atomistic simulations with a highly accurate machine learning interatomic potential. In the single crystal MPEA, we find greatly reduced anisotropy in the critically resolved shear stress between screw and edge dislocations compared to the elemental metals. In the polycrystalline MPEA, we demonstrate that thermodynamically-driven Nb segregation to the grain boundaries (GBs) and W enrichment within the grains intensifies the observed short range order (SRO). The increased GB stability due to Nb enrichment reduces the von Mises strain, resulting in higher strength than a random solid-solution MPEA. These results highlight the need to simultaneously tune GB composition and bulk SRO to tailor the mechanical properties of MPEAs.
\end{abstract}

\section{Introduction}
Multi-principal element alloys (MPEAs), colloquially also known as ``high entropy'' alloys, are alloys comprising four or more elements, usually in nearly equiatomic concentrations.\cite{gludovatzFractureresistantHighentropyAlloy2014, senkovMechanicalPropertiesNb25Mo25Ta25W252011, fengStableNanocrystallineNbMoTaW2018, liMetastableHighentropyDualphase2016a,schuhMechanicalPropertiesMicrostructure2015,juanEnhancedMechanicalProperties2015,wangExceptionalStrongFacecentered2018} They have drawn rapidly growing interest due to their exceptional mechanical properties under extreme conditions. For instance, the face-centered cubic (fcc) FeCoNiCrMn MPEA and the closely related three-component ``medium-entropy'' CrCoNi alloy have been reported to have high fracture toughness and strength, which is further enhanced at cryogenic temperatures\cite{gludovatzFractureresistantHighentropyAlloy2014,gludovatzExceptionalDamagetoleranceMediumentropy2016}. Conversely, the refractory body-centered cubic (bcc) NbMoTaW MPEA exhibits outstanding high-temperature (above 1800 K) mechanical strength\cite{senkovMechanicalPropertiesNb25Mo25Ta25W252011, fengStableNanocrystallineNbMoTaW2018}.

Despite intense research efforts, the fundamental mechanisms behind the remarkable mechanical properties of MPEAs remain under heavy debate. Solid solution strengthening, whereby the existence of multiple elements components of different atomic radii and elastic moduli impede dislocation motion, has been proposed as a key mechanism in both fcc and bcc MPEAs.\cite{varvenneTheoryStrengtheningFcc2016,senkovMicrostructureRoomTemperature2011} However, it is clear that the microstructure (e.g. nanotwinning), short-range order (SRO), phase transitions, and other effects also play significant roles\cite{ottoInfluencesTemperatureMicrostructure2013,wuNanotwinMediatedPlasticity2015,dingTunableStackingFault2018,liMetastableHighentropyDualphase2016}.

Computational simulations are an important tool to elucidate the fundamental mechanisms behind the observed strengthening in MPEAs. However, due to the high computational cost, investigations of MPEAs using density functional theory (DFT) calculations have been limited to bulk special quasi-random structures (SQSs)\cite{zhangOriginNegativeStacking2017,niuMagneticallydrivenPhaseTransformation2018,wangComputationEntropiesPhase2018}. Atomistic simulations using linear-scaling interatomic potentials (IAPs) can potentially access more complex models and longer timescales. However, classical IAPs, such as those based on the embedded atom method, are fitted mainly to elemental properties and generally perform poorly when scaled to multi-component alloys. Furthermore, classical IAPs are typically explicit parameterizations of two-body, three-body, and many-body interactions and, hence, becomes combinatorially complex for multi-element systems such as MPEAs.\cite{raoAtomisticSimulationsDislocations2017} Recently, machine learning of the potential energy surface as a function of local environment descriptors has emerged as a systematic, reproducible, automatable approach to develop IAPs (ML-IAPs) with near-DFT accuracy for elemental as well as multi-component systems.\cite{behlerGeneralizedNeuralNetworkRepresentation2007,bartokRepresentingChemicalEnvironments2013,szlachtaAccuracyTransferabilityGaussian2014,thompsonSpectralNeighborAnalysis2015,shapeevMomentTensorPotentials2016,zuoPerformanceCostAssessment2019, liQuantumaccurateSpectralNeighbor2018,dengElectrostaticSpectralNeighbor2019a,chenAccurateForceField2017} While a few ML-IAPs have been developed for MPEAs,\cite{kostiuchenkoImpactLatticeRelaxations2019} they have mainly applied to the study of phase stability of the bulk alloy.

In this work, we develop a ML-IAP for the refractory NbMoTaW alloy system using the Spectral Neighbor Analysis Potential (SNAP) approach.\cite{thompsonSpectralNeighborAnalysis2015} Using this MPEA SNAP model, we show that the Peierls stress for both screw and edge dislocation in the equi-atomic NbMoTaW MPEA are much higher than those for all the individual metals, and edge dislocations become much more important in the MPEA than that in the pure elemental bcc system. From Monte Carlo (MC)/Molecular Dynamics (MD) simulations, we find strong  evidence  of  Nb  segregation  to  the  grain boundaries (GBs) of the NbMoTaW MPEA, which in turn has a substantial effect on the observed short-range order. The observed Nb segregation to the GB leads to an enhancement in the strength of the MPEA.

\section{Results}

\subsection{NbMoTaW SNAP Model}

\begin{figure}[t]
\includegraphics[width=1.0 \textwidth]{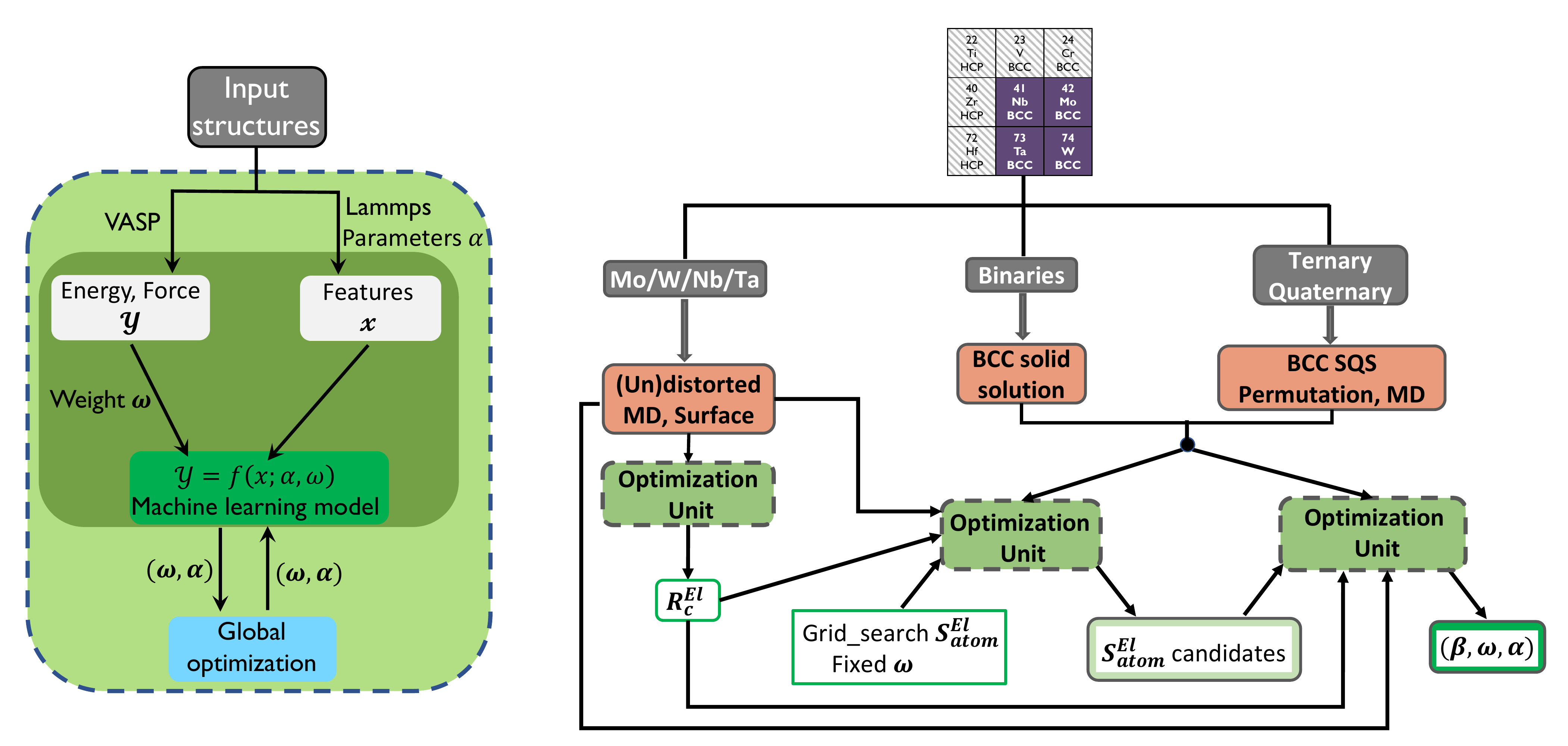}
\caption{\label{fig:workflow}Schematic for the fitting workflow for the quaternary NbMoTaW alloy SNAP model. The left panel shows one optimization unit adapted from our previous work.\cite{chenAccurateForceField2017,liQuantumaccurateSpectralNeighbor2018}. $\alpha$ are the parameters (hyperparameters) needed for bispectrum calculations, e.g. the atomic weight $S_{atom}^k$ for each element. $\beta$ are the linear SNAP model parameters. $\omega$ are data weights from different data groups.}
\end{figure}

Figure \ref{fig:workflow} shows the workflow adopted in fitting the quaternary NbMoTaW MPEA SNAP model and methodological details are provided in Methods section. Briefly, the MPEA SNAP model was fitted in three steps, as illustrated by the right panel of Figure \ref{fig:workflow} with three optimization units\cite{chenAccurateForceField2017,liQuantumaccurateSpectralNeighbor2018} from left to right. In the first step, a SNAP model was fitted for each component element, as shown in the left optimization unit of the right panel of Figure \ref{fig:workflow}. The optimized SNAP model coefficients $\beta$ are provided in Table S1, and the mean absolute error (MAE) in energies and forces are provided in Figure S1. The optimized radius cutoffs $R_c^{El}$ for Nb, Mo, Ta, W are 4.7, 4.6, 4.5, 4.5 \AA, respectively, which are slightly larger than the third nearest neighbor distance for each element. This result is in line with previous models developed for bcc elements.\cite{thompsonSpectralNeighborAnalysis2015,chenAccurateForceField2017,liQuantumaccurateSpectralNeighbor2018,zuoPerformanceCostAssessment2019} These optimized radius cutoffs were adopted for the MPEA SNAP model fittings in the next two steps. In the second step, the data weights ($\omega$) were fixed according to the number of data points for each data groups. A grid search was performed for the atomic weights ($S_{atom}^k$) by generating a series of SNAP models with different combinations of atomic weights by only running the inner loop in the optimization unit\cite{chenAccurateForceField2017,liQuantumaccurateSpectralNeighbor2018}, as shown in the middle optimization unit of the right panel of Figure \ref{fig:workflow}. In the third step (see the right optimization unit of the right panel of Figure \ref{fig:workflow}), the ten combinations of atomic weights (see Table S2) with the best accuracy in energy and force predictions was chosen to conduct a full optimization, including the optimization of the data weight in the outer loop of the optimization unit. The optimized parameters of the best model, i.e., the model with the smallest MAE in energies and forces, are provided in Table S4. The training data comprises DFT computed energies and forces for ground state structures, strained structures, surface structures, special quasi-random structures (SQS)\cite{zungerSpecialQuasirandomStructures1990} and snapshots from ab initio molecular dynamics (AIMD) simulations. A test set of structures are further generated to validate the generalizability of the fitted model, which is about 10$\%$ of that for training set. 

\begin{figure}[h]
\includegraphics[width=1.0 \textwidth]{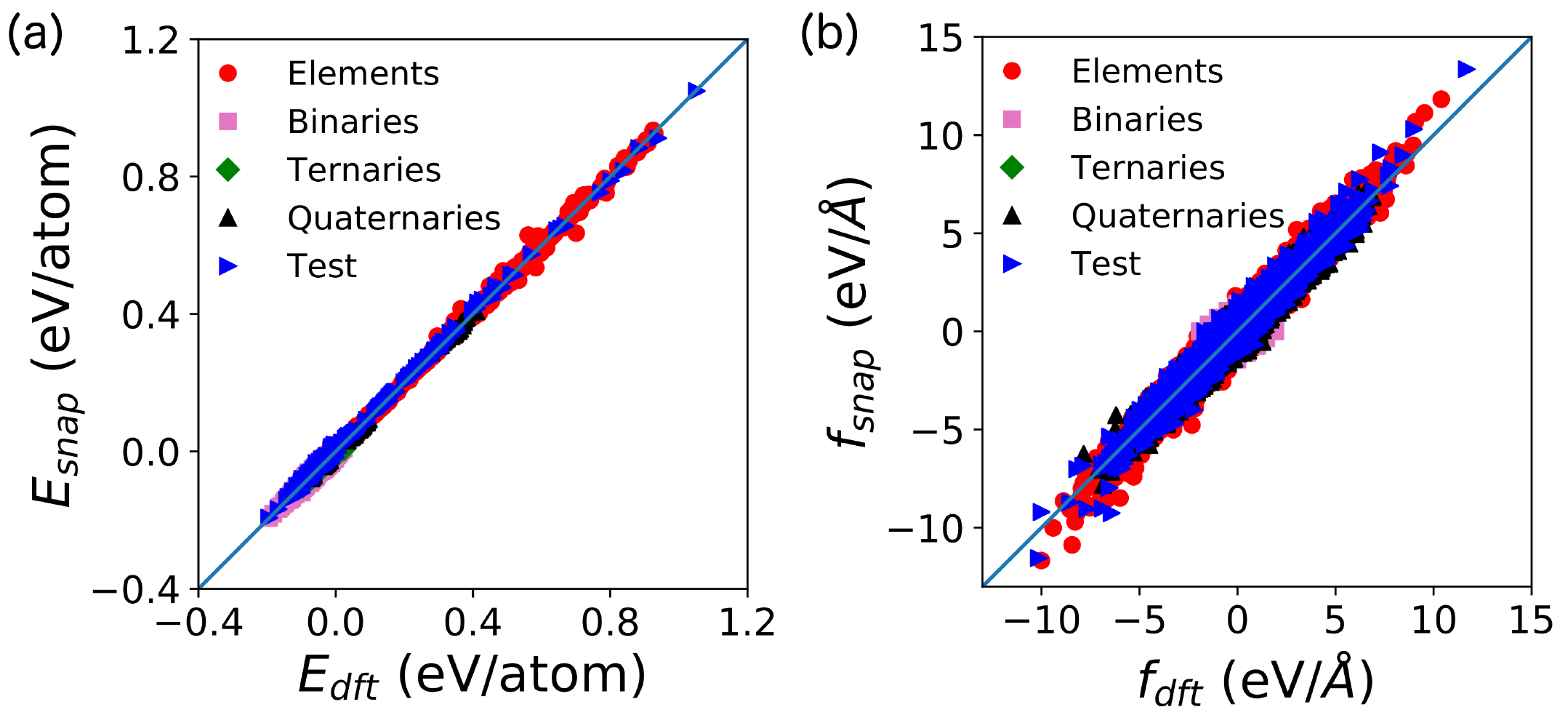}
\caption{\label{fig:quaternary}Plot of MPEA SNAP versus DFT (a) energies and (b) forces. Markers are colored according to chemical system. The overall MAEs for energies are 4.3 meV/atom and 5.1 meV/atom for the training and test set, respectively, and the overall MAEs in forces are 0.13 eV/\AA~and 0.14 eV/\AA~for the training and test set, respectively. }
\end{figure}

A comparison of the DFT and MPEA SNAP predicted energies and forces for both training and test sets is shown in Figure \ref{fig:quaternary}. The corresponding MAE values in predicted energies and forces from the SNAP model relative to DFT for the elemental, binary, ternary and quaternary systems are displayed in Table S3. An excellent fit was obtained for the MPEA SNAP model, with a unity slope in both energies and forces with respect to DFT. The overall training and test MAE for energies and forces are within 6 meV/atom and 0.15 eV/AA, respectively. More critically, this high performance is achieved consistently across all sub-chemical systems, i.e., there is no obvious bias in performance for any particular chemistry. 

\begin{center}
\begin{table}
\footnotesize
\caption{Comparison of melting points ($T_m$), elastic constants ($c_{ij}$), Voigt-Reuss-Hill\cite{hillElasticBehaviourCrystalline1952} bulk modulus (B$_{VRH}$), shear modulus (G$_{VRH}$), and Poisson's ratio ($\mu$) for bcc Nb, Mo, Ta, W, and NbMoTaW special quasi-random structure (SQS). Error percentages of the MPEA SNAP elastic properties relative to DFT values are shown in parentheses. The experimental values of B$_{VRH}$, G$_{VRH}$, and $\mu$ are derived from the experimental elastic constants. }
\label{table:elastic}
\begin{tabular}{  cccccccc }
\hline
\hline
\noalign{\smallskip}
            & {$T_m$ (K)} & $c_{11} $ (GPa)  &  $c_{12} $ (GPa)    & $c_{44} $ (GPa)     & B$_{VRH}$ (GPa)  &  G$_{VRH}$ (GPa)&$\mu$  \\
\hline
\noalign{\smallskip}
\textbf{Nb} &    &    &        &  &&&\\
Expt.  &  2750  & 247\cite{trivisonnoTemperatureDependenceElastic1973} & 135\cite{trivisonnoTemperatureDependenceElastic1973}   & 29\cite{trivisonnoTemperatureDependenceElastic1973}  &172  &38 & 0.40 \\
DFT  &  $-$  & 249 & 135   & 19  &173  &30 & 0.42 \\
SNAP &  3050 & 266(6.8\%) & 142(5.2\%)   & 20(5.3\%)  & 183(5.8\%) &32(6.7\%) & 0.42(0.0\%) \\
\hline
\noalign{\smallskip}
\textbf{Mo} &    &    &        &  &&&\\
Expt.  &  2896  & 479\cite{simmons1971single} & 165\cite{simmons1971single}   & 108\cite{simmons1971single}  &270  & 125& 0.30 \\
DFT  &  $-$  & 472 & 158   & 106  &263  & 124& 0.30 \\
SNAP &  3420  & 435($-7.8\%$) & 169(7.0\%)   & 96($-9.4\%$)  & 258($-1.9\%$) &110($-11.3\%$) & 0.31(3.3\%) \\
\hline
\noalign{\smallskip}
\textbf{Ta} &    &    &        &  &&&\\
Expt.  &  3290  & 266\cite{featherstonElasticConstantsTantalum1963} & 158\cite{featherstonElasticConstantsTantalum1963}  & 87\cite{featherstonElasticConstantsTantalum1963}  &194  &72 & 0.34 \\
DFT  &  $-$  & 264 & 161   & 74  &195  &64 & 0.35 \\
SNAP &  3540  & 257($-2.7\%$) & 161(0.0\%)   & 67($-9.5\%$)  & 193($-1.0\%$) &59($-7.8\%$) & 0.36(2.9\%) \\
\hline
\noalign{\smallskip}
\textbf{W} &    &    &        &  &&&\\
Expt.  &  3695  & 533\cite{featherstonElasticConstantsTantalum1963} & 205\cite{featherstonElasticConstantsTantalum1963}  & 163\cite{featherstonElasticConstantsTantalum1963}  &314  &163 & 0.28 \\
DFT  &  $-$  & 511 & 200  & 142  &304  &147 & 0.29 \\
SNAP &  4060  & 560(9.6\%) & 218(9.0\%)   & 154(8.5\%)  & 332(9.2\%) &160(8.8\%) & 0.29(0.0\%) \\
\hline
\noalign{\smallskip}
\multicolumn{2}{c}{\textbf{NbMoTaW SQS}}     &    &        &  &&&\\
DFT  &  $-$  & 377 & 160   & 69  &233  &83 & 0.34 \\
SNAP &  3410  & 399(5.8\%) & 166(3.8\%)   & 80(15.9\%)  & 243(4.3\%) &94(13.3\%) & 0.33($-2.9\%$) \\

\hline
\hline
\end{tabular}

\end{table}
\end{center}

The MPEA SNAP was further validated by computing various properties of the elements and multi-component systems, as presented in  Table \ref{table:elastic}. While the MPEA SNAP model systematically overestimates the melting points for all elements, they are still in qualitative agreement with the experimental values. The elastic moduli predicted by the MPEA SNAP model are in extremely good agreement with the DFT for all four elemental systems, with errors of less than 10\% except for the shear modulus of Mo ($-11.3\%$). The MPEA SNAP also performs very well on the NbMoTaW special quasi-random structure (SQS), except for a slight overestimate of $c_{44}$ and the corresponding shear modulus. It should be noted that only strained elemental structures, and not strained SQS structures, were included in the training data. Therefore, this excellent performance on the quaternary NbMoTaW SQS is an important validation test for the generalizability of the MPEA SNAP model. 

\subsection{Generalized stacking fault energies}

\begin{figure}[t]
\includegraphics[width=1.0 \textwidth]{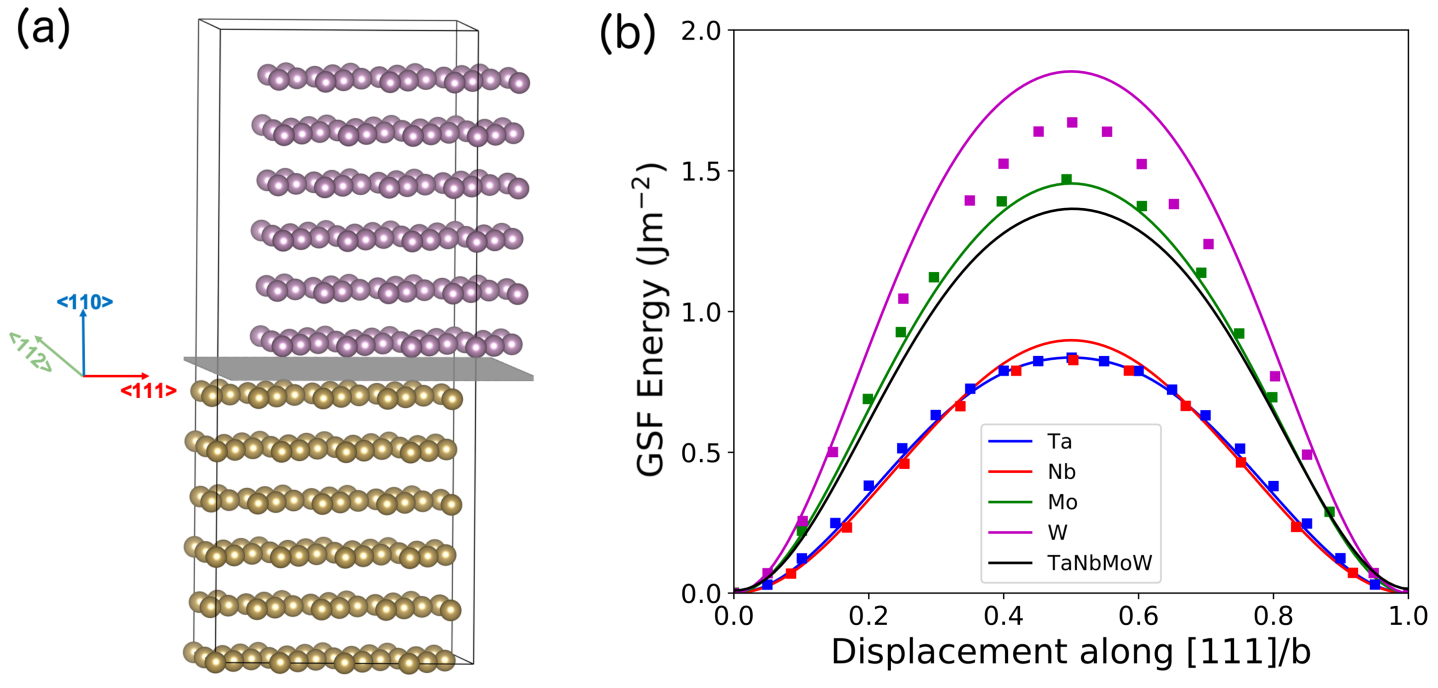}
\caption{\label{fig:GSF}(a) Schematic view of the shifts along $<$111$>$ of $\{110\}$ plane. (b) Comparison between MPEA (lines) and DFT (square markers) energies of the 1/2 $<111>\{110\}$ generalized stacking fault (GSF) of the elements and SQS NbMoTaW structure. The DFT GSF energies for Ta, Nb, W and Mo are obtained from references \citenum{mishinAngulardependentInteratomicPotential2006}, \citenum{fellingerForcematchedEmbeddedatomMethod2010}, \citenum{bonnyManybodyCentralForce2014}, and \citenum{frederiksenDensityFunctionalTheory2003}, respectively.}
\end{figure}

Metastable stacking faults play a critical role in the dissociation of dislocations in bcc metals.\cite{poPhenomenologicalDislocationMobility2016,gordonScrewDislocationMobility2010}. The $\gamma$ surfaces represent energies of generalized stacking faults (GSFs), formed by shifting two halves of a crystal relative to each other along a crystallographic plane\cite{vitekIntrinsicStackingFaults1968}. The MPEA SNAP model was used to compute a section along the $\{110\}$ $\gamma$ surface in the $<$111$>$ direction (see Fig \ref{fig:GSF}a) for all four elemental bcc metals and the NbMoTaW SQS. Fig \ref{fig:GSF}b shows the comparison between the MPEA SNAP results and previous DFT studies\cite{mishinAngulardependentInteratomicPotential2006,fellingerForcematchedEmbeddedatomMethod2010,bonnyManybodyCentralForce2014, frederiksenDensityFunctionalTheory2003}. We can see that the overall agreement between SNAP and DFT results for all four elemental systems is excellent, with only a slight overestimation for the W. No local minima that would indicate the existence of metastable stacking faults was found in the elemental metals and the NbMoTaW SQS. The stacking fault energies of the NbMoTaW SQS are smaller than that of W and Mo, and much larger than those of Ta and Nb.

\subsection{Dislocation core structure}

\begin{figure}[t]
\includegraphics[width=1.0 \textwidth]{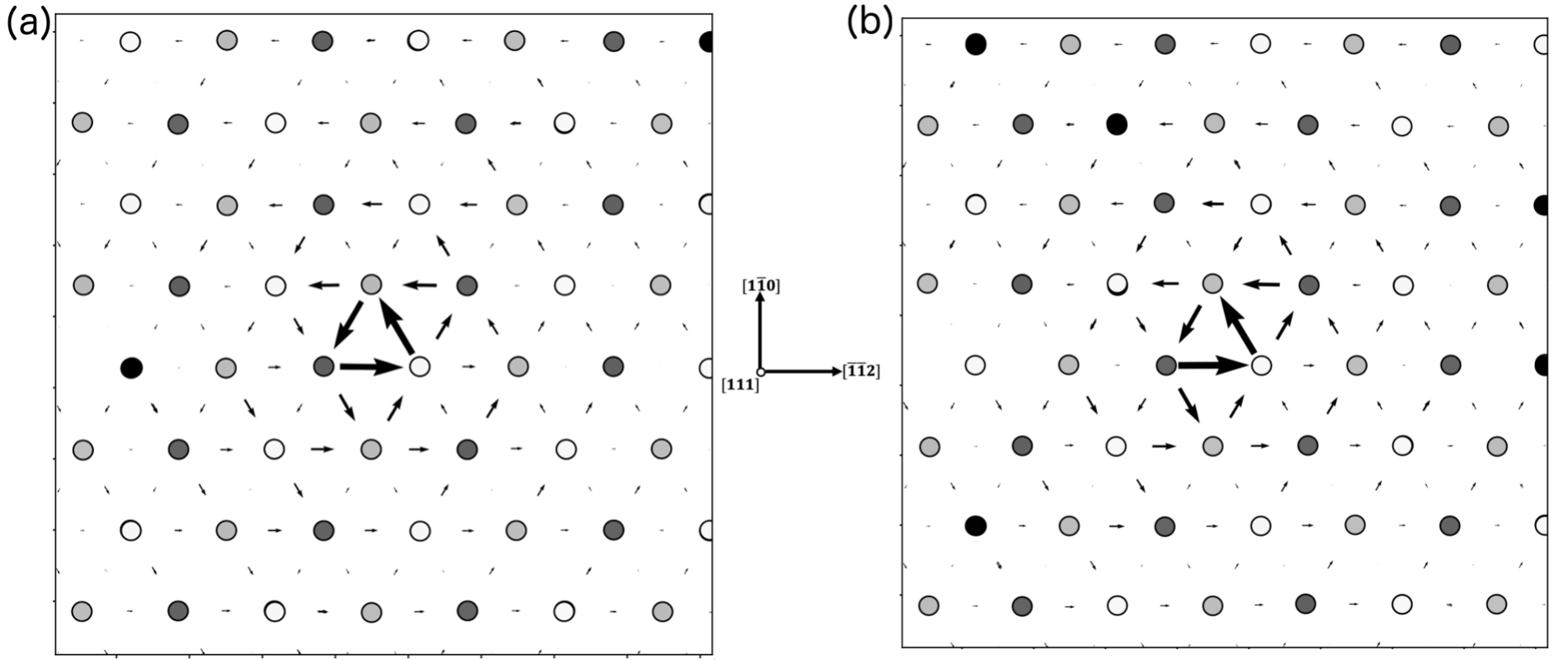}
\caption{\label{fig:core}Differential-displacement maps for the 1/2 $<$111$>$ screw dislocation core structure of the NbMoTaW SQS MPEA at (a) $0b$ and (b) $4b$ along the dislocation line, where $b$ is the length of the Burgers vector. The color depth indicates the layer of the atoms. This core structure was obtained by placing a 1/2$<$111$>$ screw dislocation at the center of a cylinder supercell with a radius of 10 nm and carrying out full relaxation using the MPEA SNAP model.}
\end{figure}

One of the important characteristics of the screw dislocation in bcc metals is the core structure. Two types of 1/2$<$111$>$ screw dislocation core structures have been reported in previous calculations for bcc metals\cite{raoAtomisticSimulations1112001,duesberyPlasticAnisotropyTransition1998,ismail-beigiInitioStudyScrew2000, frederiksenDensityFunctionalTheory2003, woodwardFlexibleInitioBoundary2002}, the degenerate core and the non-degenerate (or symmetric) core. A major discrepancy between \textit{ab initio} methods and classical force fields is that the former predicts the non-degenerate configuration to be the core structure of the 1/2$<$111$>$ screw dislocation in bcc metals\cite{ismail-beigiInitioStudyScrew2000, frederiksenDensityFunctionalTheory2003, woodwardFlexibleInitioBoundary2002}, while the latter generally finds the degenerate core.\cite{raoAtomisticSimulations1112001,duesberyPlasticAnisotropyTransition1998} The MPEA SNAP model accurately predicts the non-degenerate core structure for the 1/2$<$111$>$ screw dislocation for all the bcc elements, consistent with DFT. Fig.\ref{fig:core} shows classical differential displacement\cite{vitekCoreStructure1111970} plots at $0b$ and $4b$ along the dislocation line ($b$ is the length of the Burgers vector) for the 1/2 $<$111$>$ screw dislocation of the NbMoTaW SQS MPEA. It may be observed that there is substantial variation in the core structure in the MPEA due to the different local chemical environments. The core structure is compact at $0b$, but shows more non-compact characteristics at $4b$. Similar variations of the core structures in MPEA due to compositional fluctuations have been reported recently.\cite{raoModelingSolutionHardening2019, yinInitioModelingRole2019} Nevertheless, sampling hundreds of local environments in the SQS cell indicate that the majority of local environments exhibit a compact core as shown in Fig.\ref{fig:core}a, and only a very few local environments exhibit more non-compact characteristics as illustrated in Fig.\ref{fig:core}b. These findings obtained from the SNAP calculations are consistent with the recent DFT studies in the same quaternary system.\cite{yinInitioModelingRole2019}  

\subsection{Critical resolved shear stress of screw and edge dislocations}

\begin{center}
\large
\begin{table}
\caption{Calculated critical resolved shear stress (in MPa) of 1/2 $<$111$>$ dislocations in bcc elemental and MPEA systems, and compared with previous computational (Prev. Comp.) and experimental (Expt.) values. Previous computational results from refs \citenum{fellingerForcematchedEmbeddedatomMethod2010}, \citenum{woodwardFlexibleInitioBoundary2002} and \citenum{tianMovementScrewDislocations2004} were obtained using the embedded atom method, ab initio calculations and the Ackland potential, respectively.}
\label{table:CRSS}
\begin{tabular}{  cccccc }
\hline
\hline
\noalign{\smallskip}
           Method(dislocation type) & {\textbf{Nb}} & {\textbf{Mo}}  &  {\textbf{Ta}}    & {\textbf{W}}     & {\textbf{NbMoTaW SQS}}   \\
\hline

Expt. (screw)  &   415\cite{kamimuraExperimentalEvaluationPeierls2013} &730\cite{kamimuraExperimentalEvaluationPeierls2013}& 340\cite{kamimuraExperimentalEvaluationPeierls2013} & 960\cite{kamimuraExperimentalEvaluationPeierls2013} & $-$\\
Prev. Comp. (screw) &   1339\cite{fellingerForcematchedEmbeddedatomMethod2010} &2363\cite{woodwardFlexibleInitioBoundary2002}& 1568\cite{woodwardFlexibleInitioBoundary2002} & 3509\cite{tianMovementScrewDislocations2004} & $-$\\
SNAP (screw)  &   889 &1376& 912 & 1686 & 1620$\pm$637\\
SNAP (edge)  &   29 &76& 41 & 56 & 320$\pm$113\\

\hline
\hline
\end{tabular}

\end{table}
\end{center}

The critical resolved shear stress (CRSS) to move dislocations is closely related to the strength of the materials. Table \ref{table:CRSS} shows the calculated CRSS for screw and edge dislocations for all four elements and the NbMoTaW SQS, together with previous experimental and computed values for the screw dislocation where available. The calculated CRSSs from atomistic simulations are typically much larger than experimentally measured values, a well-known discrepancy in bcc crystals attributed to the quantum effect\cite{provilleQuantumEffectThermally2012}. It may be observed that the MPEA SNAP CRSS for screw dislocations are substantially closer to the experimental values. More importantly, the qualitative trends in the CRSS for screw dislocations is successfully reproduced, i.e., W has the largest CRSS, followed by Mo, with Ta and Nb having much smaller CRSS. The MPEA SNAP CRSS of edge dislocations in the elements are an order of magnitude smaller than the CRSS of screw dislocations, consistent with previous studies.\cite{wangHydrogeninducedChangeCore2013,kangSingularOrientationsFaceted2012} This leads to the well-known large screw/edge anisotropy in apparent mobility and the dominance of screw dislocations in the deformation of bcc metals\cite{caillardKineticsDislocationsPure2010}. For NbMoTaW SQS MPEA system, Fig. S3 shows the distribution of the calculated CRSS for both screw and edge dislocations for different local chemical environments. It may be observed that the local environment can have a substantial effect on the CRSS. The average and standard deviation of the CRSS for the screw and edge dislocation in the SQS MPEA are reported in Table \ref{table:CRSS}. Genereally, the MPEA SNAP model predicts a very high CRSS for screw dislocation in NbMoTaW SQS, much larger than that of Nb, Ta, Mo and comparable with that of W. The most interesting observation, however, is that the MPEA SNAP CRSS for the edge dislocation in the NbMoTaW SQS is also much higher than those of the elemental components and is about $\sim$20\% of the CRSS of the screw dislocation. This greatly-reduced anisotropy between screw and edge mobility suggests that the edge dislocation may play a more important role in the deformation of the bcc MPEA compared to in bcc elements.

\subsection{Segregation and short range order}

\begin{figure}[t]
\includegraphics[width=1.0 \textwidth]{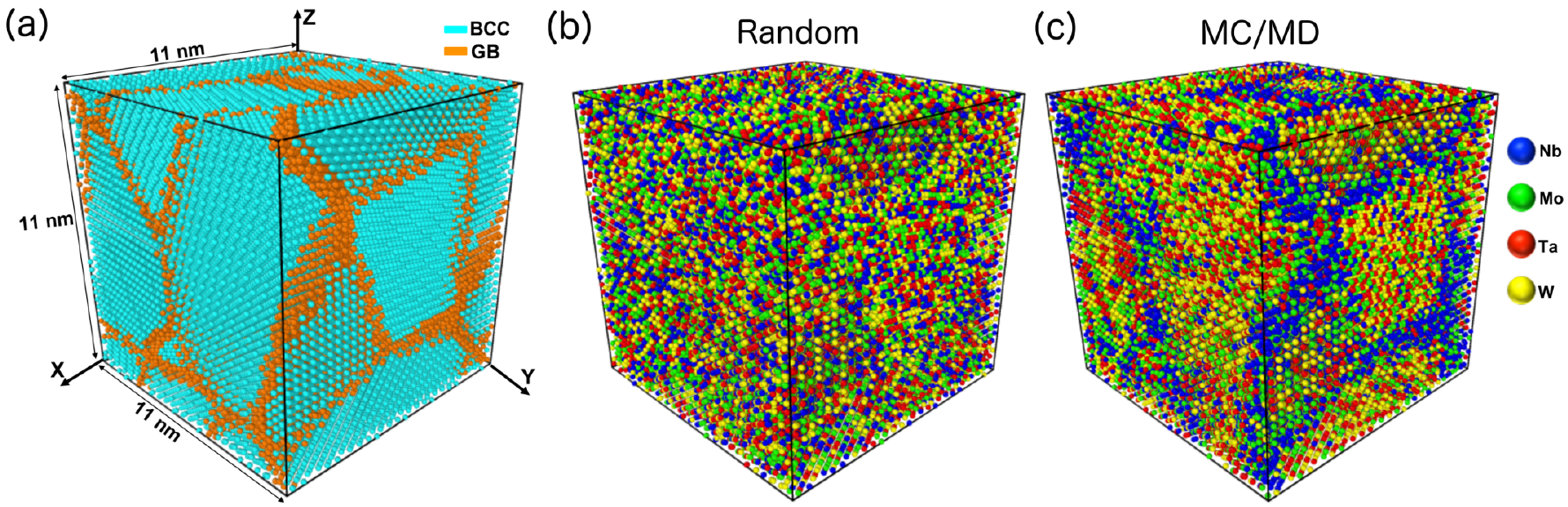}
\caption{\label{fig:segre}(a) A polycrystalline model for the quaternary NbMoTaW MPEA with atoms colored according to the common neighbor analysis algorithm\cite{honeycuttMolecularDynamicsStudy1987} in OVITO\cite{stukowskiVisualizationAnalysisAtomistic2009} to identify different structure types (cyan: bulk bcc; orange: grain boundary). (b) The same polycrystalline model after random initialization with equi-molar quantities of Nb, Mo, W, and Ta. Atoms are colored by element. (c) Snapshot of polycrystalline model after hybrid Monte Carlo/MD simulations. Clear segregation of Nb to the grain boundaries can be observed.}
\end{figure}

\begin{figure}[t]
\includegraphics[width=1.0 \textwidth]{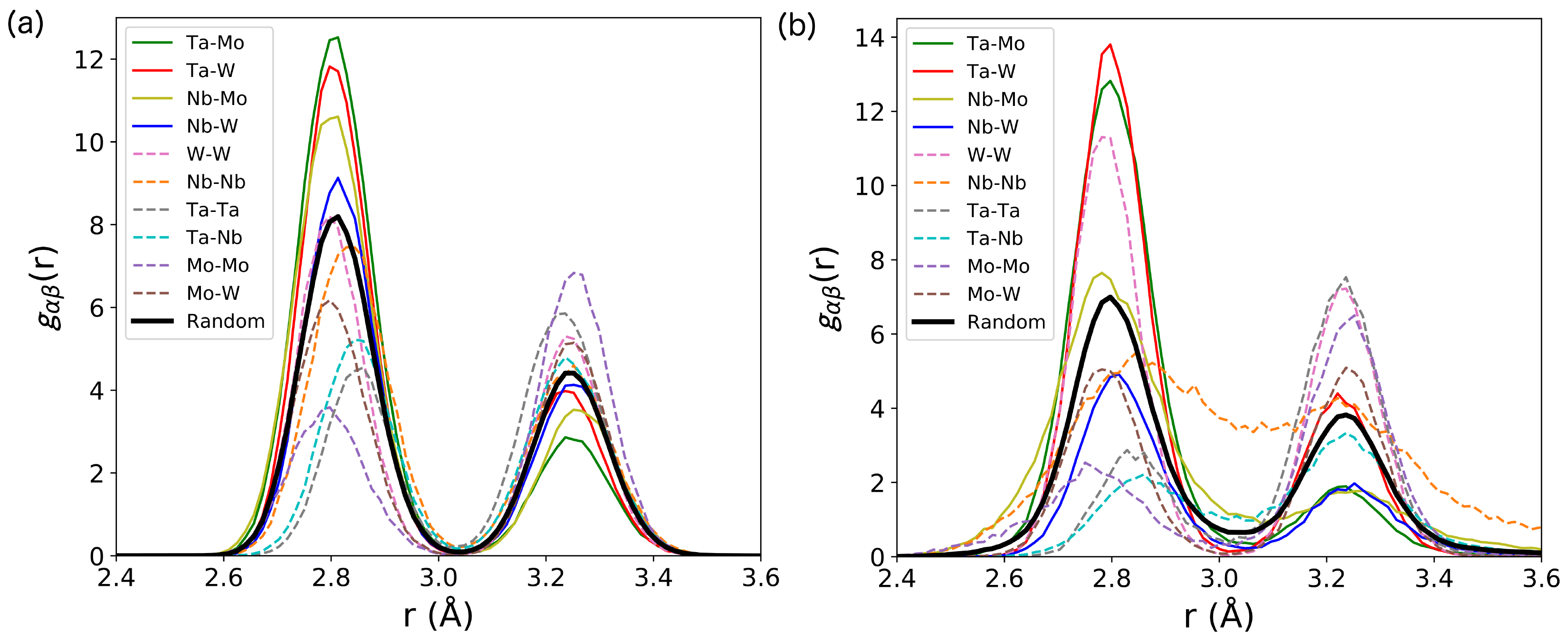}
\caption{\label{fig:rdf}Partial radial distribution function $\mathrm{g}_{\alpha\beta}$ at T = 300 K for (a) a single-crystal NbMoTaW MPEA and (b) a polycrystalline  NbMoTaW MPEA.}
\end{figure}

\begin{center}
\large
\begin{table}
\caption{Pairwise chemical short-range order parameters $\alpha^1$(see Methods) for the NbMoTaW MPEA single crystal and polycrystal after MC/MD equilibration at room temperature. }
\label{table:srop}
\begin{tabular}{  ccc }
\hline
\hline
\noalign{\smallskip}
           Pairs & Single crystal& Polycrystal  \\
\hline
Ta-Mo  &   $-0.51$&$-0.71$\\
Ta-W  &   $-0.34$&$-0.61$\\
Nb-Mo  &   $-0.33$&$-0.35$\\
Nb-W  &   $-0.05$&$0.34$\\
W-W  &   $-0.04$&$0.09$\\
Nb-Nb  &   $-0.01$&$0.10$\\
Ta-Ta  &   $-0.16$&$-0.21$\\
Ta-Nb  &   $0.36$&$0.62$\\
Mo-Mo  &   $-0.19$&$-0.19$\\
Mo-W  &   $0.28$&$0.36$\\

\hline
\hline
\end{tabular}
\end{table}
\end{center}

\begin{center}
\large
\begin{table}
\caption{The atomic percentages of each element at the grain boundary and in the BCC bulk region for the initial random structure and after MC/MD simulations. }
\label{table:seg}
\begin{tabular}{  ccccc|cccc }
\hline
\hline
\noalign{\smallskip}
           Region & \multicolumn{4}{c} {\textbf{Grain Boundary}} & \multicolumn{4}{c} {\textbf{BCC bulk}}   \\
\hline
Element  &   Nb &Mo& Ta & W &Nb&Mo&Ta&W\\
Random  &   24.3 &25.5& 24.2 & 26.0 &25.0&25.0&25.0&25.0\\
MC/MD  &   57.7 &26.7& 13.6 & 2.0 &15.5&24.6&28.0&31.9\\
\hline
\hline
\end{tabular}
\end{table}
\end{center}

The validated MPEA SNAP model was applied to long-time, large-scale simulations of both single crystal and polycrystalline models of the NbMoTaW MPEA. The single-crystal and polycrystal models were constructed using supercells of dimensions $15.5\times 15.5\times 15.5$ nm ($48\times 48\times 48$ conventional cell) and $11\times 11\times 11$ nm (Fig. \ref{fig:segre} a), respectively, with a randomized elemental distribution with equi-atomic quantities, i.e., 25\% each, of Nb, Mo, W, and Ta (see Fig. \ref{fig:segre} b). Hybrid MC/MD simulations were then performed to obtain low energy microstructures for the quaternary NbMoTaW MPEA at room temperature (see Methods for details). 

One important property that can be analyzed is the structural characteristics, such as pair correlation functions. For the single-crystal MPEA, the partial pair correlation functions are plotted in Fig. \ref{fig:rdf}a for both the random structure and the structure after equilibration in the MC/MD simulations. The dominant nearest-neighbor correlations in the structure after MC/MD equilibration are between elements in different groups in the periodic table, with Ta-Mo being the highest, followed by Ta-W, Nb-Mo and Nb-W; the correlations between elements within the same group (Ta-Nb and Mo-W) are much lower. This is consistent with the enthalpies of pairwise interactions (see Table S5). The non-uniform correlations indicate the existence of local chemical order (LCO) in the structure at room temperature. The computed pairwise multi-component short-range order (SRO) parameters\cite{fontaineNumberIndependentPaircorrelation1971,liStrengtheningMultiprincipalElement2019} (see Methods) are presented in Table \ref{table:srop}. It was found that three inter-group elemental pairs - Ta-Mo, Ta-W, Nb-Mo -  have large negative SRO parameters, indicating attractive interactions between these elements. Large positive SRO parameters are observed between elements within the same group. Our findings about SRO in single-crystal NbMoTaW MPEA are also consistent with previous studies\cite{kormannLongrangedInteractionsBcc2017,kostiuchenkoImpactLatticeRelaxations2019}, showing that Ta-Mo pairs have the most dominant SRO followed by Ta-W, Nb-Mo, and Nb-W, as such a B2(Mo,W;Ta,Nb) phase is observed at intermediate temperature.

For the MPEA polycrystal, Fig. \ref{fig:segre}c shows a snapshot of the polycrystalline model after equilibration in the MC/MD simulations. Clear segregation of Nb (blue atoms) to the GBs can be observed, while there is evidence of an enrichment of W in the bulk. Table \ref{table:seg} provides the atomic percentages for each element in the GBs and bulk before and after the MC/MD simulations. Starting with an initial equal distribution of approximately 25\% for all elements in both GBs and bulk, the percentage of Nb in the GBs increases to $\sim 57.7 \%$, while the percentage for W and Ta decreases to $\sim2\%$ and $\sim 13.6\%$, respectively.  Correspondingly, the percentage for W and Ta in the bulk regions increase to $\sim32\%$ and $\sim 28\%$, respectively, while that for Nb decreases to $\sim 16\%$. The corresponding partial radial distribution functions are also plotted for this polycrystalline model after equilibration in the MC/MD simulations, as shown in Fig. \ref{fig:rdf}b. We can clearly see a large decrease for the nearest neighbor peak of Nb-W (blue curve) compared to the single-crystal model due to the segregation effects. In addition, the peaks for the Nb-Nb pair are broadened due to the more disorder characteristics of Nb-Nb pairs segregated into the GBs. The segregation of W into the grain interior also leads to a large increase for the nearest neighbor peak of W-W. The calculated SRO parameters (see Table \ref{table:srop}) also indicate very different interactions between the polycrystal and single crystal MPEAs. For example, $\alpha^1_{Nb-W}$ changes from a small negative value (-0.05) to a large positive value (0.34), due to the tendency of Nb and W to segregation into the GB and bulk regions, respectively.

\begin{figure}[t]
\includegraphics[width=1.0 \textwidth]{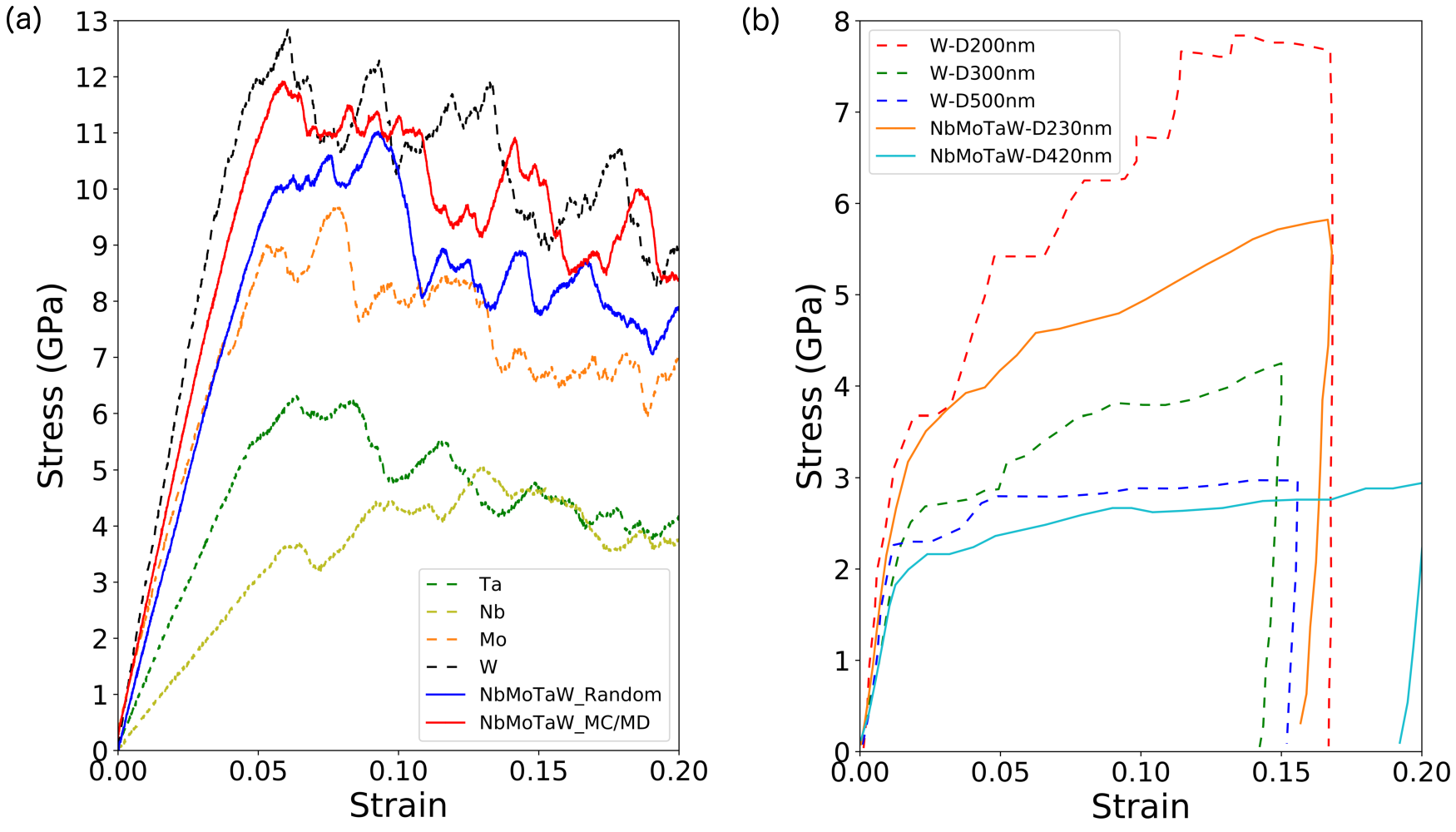}
\caption{\label{fig:stress} (a) SNAP predicted uniaxial compressive stress-strain behaviour of polycrystals of four elemental systems (dashed line) and the quaternary system (solid line) with atoms randomly distributed and segregated after MC/MC simulations. (b) Experimental compressive stress-strain behaviour of nanopillars of W (dashed line) and the quaternary system (solid line) with different diameters (D) extracted from references \citenum{schneiderCorrelationCriticalTemperature2009} and \citenum{zouSizedependentPlasticityNb25Mo25Ta25W252014}, respectively.}
\end{figure}

MD simulations using the MPEA SNAP model were performed to generate uniaxial compressive stress-strain responses of nanocrystalline models of the elements as well as the random NbMoTaW MPEA and the equilibrated NbMoTaW MPEA after MC/MD simulations, as shown in Fig. \ref{fig:stress}a. Among the elements, W has the highest strength, followed by Mo, and Ta and Nb being much weaker. The random MPEA has a strength that is substantially higher than that of Mo, Nb and Ta. Most interestingly, the MC/MD-equilibrated NbMoTaW MPEA exhibits a substantially higher strength than the random solid solution MPEA and close to that of W, the strongest elemental component. These results are consistent with previous experimental measurements\cite{schneiderCorrelationCriticalTemperature2009,zouSizedependentPlasticityNb25Mo25Ta25W252014} , which found that nanopillars of the MPEA has comparable compressive stress-strain curves with those of W at similar diameters (Fig. \ref{fig:stress}b). 

\section{Discussion}

We have developed a highly-accurate spectral neighbor analysis potential (SNAP) for the four-component Nb-Mo-Ta-W system and applied it in large-scale, long-time simulations of both single crystal and polycrystal NbMoTaW MPEAs. The accuracy of the MPEA SNAP model has been thoroughly evaluated based on not just accuracy in energy and force predictions, but also in the reproduction of key mechanical properties such as the elastic constants, dislocation core structure and critically-resolved shear stress (CRSS). 

In the single crystal MPEA, we find strong evidence of a reduced screw/edge anisotropy in the calculated CRSS. This finding supports recent observations that edge dislocations may play a far more important role in MPEAs as compared to bcc elements.\cite{mompiouConventionalVsHarmonicstructured2018,marescaMechanisticOriginHigh2020}. For example, \citet{mompiouConventionalVsHarmonicstructured2018} have found that edge dislocations are sluggish at room temperature in the Ti$_{50}$Zr$_{25}$Nb$_{25}$ alloy, indicating the comparable role in strength of edge to screw dislocations. \citet{marescaMechanisticOriginHigh2020} have also proposed that the random field of solutes in the high-concentration alloys has been found to create large energy barriers for thermally-activated edge glide, and established a theory of strengthening of edge dislocations in BCC alloys.

Large-scale, long-time simulations using the MPEA SNAP model have also provided critical new insights into the interplay between segregation, short range order and mechanical properties of the \textit{polycrystalline} NbMoTaW MPEA for the first time. First, it was found that there is a clear tendency for Nb to segregate to the GB, accompanied by enrichment of W in the bulk. Similar elemental segregation to GBs have also been observed in FeMnNiCoCr MPEA after aging heat treatment in a recent experiment\cite{liSegregationdrivenGrainBoundary2019}. This effect can be explained by considering the relative GB energies of the different elements. The current authors have previously developed a large public database of GB energies for the elemental metals using DFT computations\cite{zhengGrainBoundaryProperties2019}. As shown in Fig. S2, Nb has the lowest GB energy and W has the highest among the four component elements. Hence, Nb segregation to the GB region and W enrichment in the bulk is driven by a thermodynamic driving force to lower the GB energies.

\begin{figure}[t]
\includegraphics[width=1.0 \textwidth]{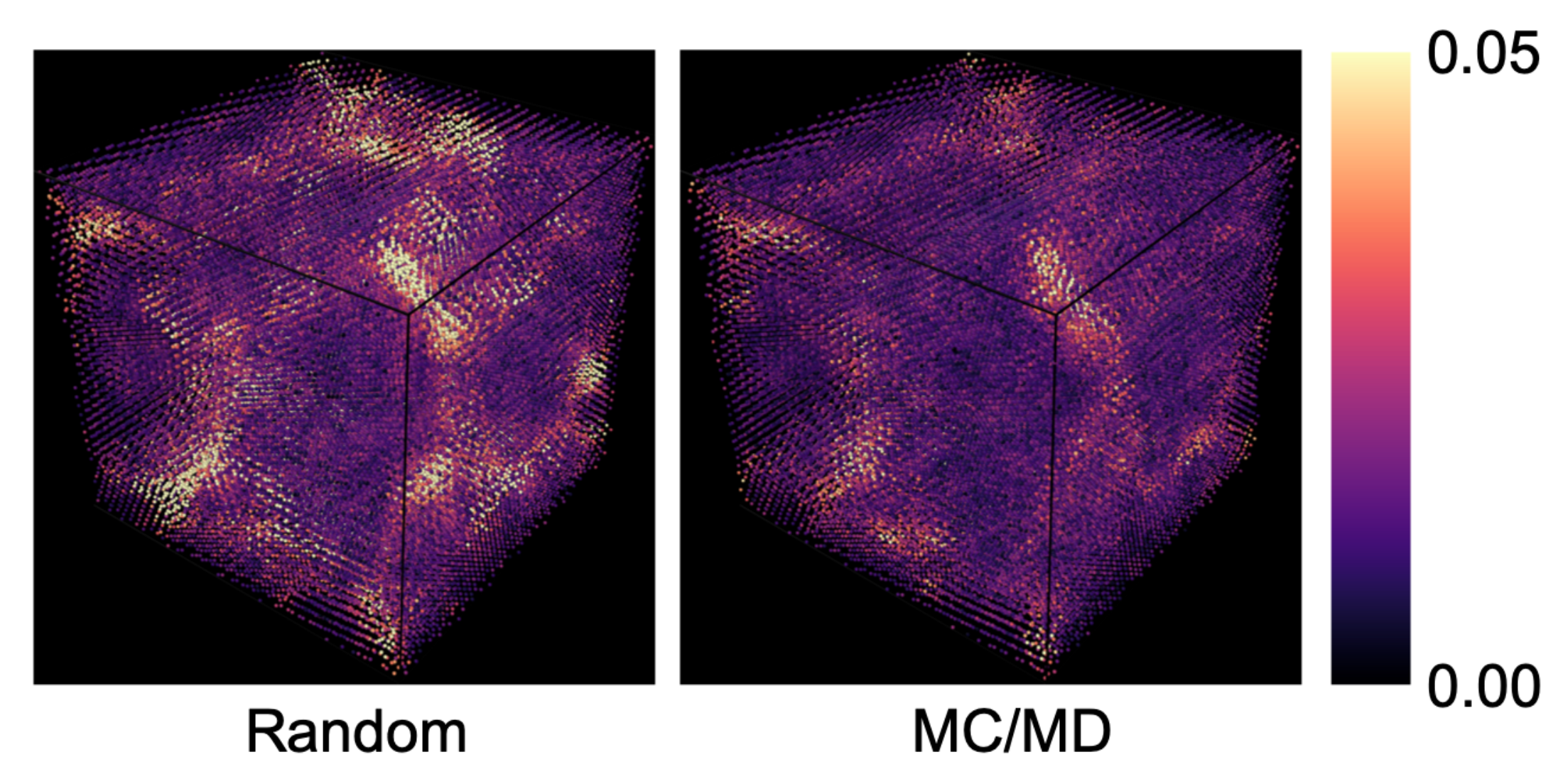}
\caption{\label{fig:ls}Atomic Von Mises strain maps of both the random NbMoTaW polycrystal and the polycrystal after equilibration in the MC/MD simulations at an uniaxial compressive strain of 3.0$\%$.}
\end{figure}

In turn, Nb segregation has substantial effect on the observed SRO in the NbMoTaW MPEA. As can be seen from Table \ref{table:srop}, the Nb-W SRO parameter changes from a small attractive interaction in the single crystal to a strong repulsive interaction in the polycrystal due to Nb segregation to the GB and W to the bulk. The SRO parameters of other pairs of elements are also intensified in magnitude. An increase in SRO has been found to lead to increased barriers to dislocation motion, leading to greater strength in the fcc NiCoCr MPEA\cite{liStrengtheningMultiprincipalElement2019}. Indeed, a similar effect is observed in the polycrystalline MPEA, where the equilibrated NbMoTaW MPEA with SRO exhibiting substantially higher strength than the random solid solution NbMoTaW MPEA, with a strength approaching that of W (Fig. \ref{fig:stress}a). The von Mises strain distribution\cite{falkDynamicsViscoplasticDeformation1998,shimizuTheoryShearBanding2007} at a low applied strain of 3.0$\%$ is plotted in Fig. \ref{fig:ls}. It can be observed that the von Mises strain distribution is localized in the GB region for both the random solid solution and the equilibrated MPEA. However, the MC/MD equilibrated polycrystal with Nb-rich GBs shows much smaller von Mises strains than the random solid solution. Similar GB stability-induced strengthening has also been observed experimentally in Ni-Mo nanograined crystals\cite{huGrainBoundaryStability2017}. 

To conclude, this work has highlighted that accurate treatment of MPEA chemistry at the inter- and intra-granular is necessary to reveal the subtle interactions between segregation and SRO, and their consequent effect on mechanical properties. Interestingly, Nb enrichment of the GB coupled with intensification of the SRO are predicted to enhancement of the strength of the NbMoTaW MPEA over a random solid solution. These findings point to the potential for leveraging on both composition and processing modification to tune the GB composition and bulk SRO to tailor the mechanical properties of MPEAs.

\section{Methods}

\subsection{Bispectrum and SNAP formalism}

The Spectral Neighbor Analysis Potential (SNAP) model expresses the energies and forces of a collection of atoms as a function of the coefficients of the bispectrum of the atomic neighbor density function\cite{bartokGaussianApproximationPotentials2010}. The atomic neighbor density function is given by:
\begin{equation}
\rho_i(\textbf{r}) = \delta(\textbf{r}) +\\ \sum_{r_{ik}<R_c}f_c(r_{ik})S_{atom}^k\delta(\textbf{r}-\textbf{r}_{ik}).\label{eq1}
\end{equation}
where $\delta(\textbf{r}-\textbf{r}_{ik})$ is the Dirac delta function centered at each neighboring site $k$, the cutoff function $f_c$ ensures a smooth decay for the neighbor atomic density to zero at the cutoff radius $R_c$, and the atomic weights $S_{atom}^k$ distinguishes between different atom types. This atomic density function can be expanded as a generalized Fourier series in the 4D hyperspherical harmonics $U_{m, m^{\prime}}^{j}$ as follows:
\begin{equation}
    \rho_{i}(\textbf{r}) = \sum_{j=0}^{\infty} \sum_{m, m^{\prime}=-j}^{j} u_{m, m^{\prime}}^{j} U_{m, m^{\prime}}^{j}.
\end{equation}
where $u_{m, m^{\prime}}^{j}$ are the coefficients. The bispectrum coefficients are then given by:
\begin{eqnarray}
\textbf{\it B}_{j_1,j_2,j} = \sum_{m_1,m_1^{'} = -j_1}^{j_1}\sum_{m_2,m_2^{'} = -j_2}^{j_2}\sum_{m,m^{'} = -j}^j(u_{m,m^{'}}^j)^*\cdot C_{j_1m_1j_2m_2}^{jm}\times C_{j_1m_1^{'}j_2m_2^{'}}^{jm^{'}} u_{m_1^{'},m_1}^{j_1}u_{m_2^{'},m_2}^{j_2},\label{eq3}
\end{eqnarray}
where $C_{j_1m_1^{'}j_2m_2^{'}}^{jm^{'}}$ are the Clebsch-Gordon coupling coefficients.

In the linear SNAP formalism,\cite{thompsonSpectralNeighborAnalysis2015,chenAccurateForceField2017} the energy $E$ and force on atom $j$ $\mathbf{f}_j$ are expressed as a linear function of the $K$ projected bispectrum components $B_{k}$ and their derivatives, as follows:
\begin{eqnarray}
E_{SNAP} =& \sum_{\alpha}\left(\beta_{\alpha,0}N_\alpha + \sum^{K}_{k=1}\beta_{\alpha,k}\sum^{N_\alpha}_{i=1}B_{k,i}\right)\\
\mathbf{f}_{j,SNAP} =& - \sum_\alpha\sum^{K}_{k=1}\beta_{\alpha,k}\sum^{N_\alpha}_{i=1}\frac{\partial B_{k,i}}{\partial \mathbf{r}_j}.
\end{eqnarray}
where $\alpha$ is the chemical identity of atoms, $N_\alpha$ is the total number of $\alpha$ atoms in the system, and $\beta_{\alpha,k}$ are the coefficients in the linear SNAP model for type $\alpha$ atoms. 

The key hyperparameters influencing model performance are the cutoff radius $R_c$ for bispectrum computation, atomic weight $S_{atom}^k$ for element $k$ (Nb, Mo, Ta or W)  and the order of the bispectrum coefficients $j_{max}$. In this work, the LAMMPS package\cite{plimptonFastParallelAlgorithms1995} was used to calculate the bispectrum coefficients (the features) for all the training structures.\cite{thompsonSpectralNeighborAnalysis2015} An order of three for the bispectrum coefficients ($j_{max} =3$) was used, consistent with previous works\cite{bartokGaussianApproximationPotentials2010,thompsonSpectralNeighborAnalysis2015,chenAccurateForceField2017,liQuantumaccurateSpectralNeighbor2018}. The cutoff radius $R_c$ and atomic weight $S_{atom}^k$ were optimized during the training of the model.
 
\subsection{Training Data generation}

One critical factor for developing an effective and robust potential is a diverse training data encompassing a good range of atomic local environments. For a quaternary potential, the training data should include the elemental, binary, ternary and quaternary systems. The detailed structure generation for each system is provided as follows.
 \begin{enumerate}
    \item Elemental systems (Nb, Mo, Ta, W)
    \begin{enumerate}
        \item Undistorted ground state structure for the element;
        \item Distorted structures constructed by applying strains of $-10\%$ to 10$\%$ at 1$\%$ intervals to the bulk conventional cell of the element in six different modes\cite{dejongChartingCompleteElastic2015};
        \item Surface structures of elemental system\cite{tranSurfaceEnergiesElemental2016,crystalium}; 
        \item Snapshots from $NVT$ \textit{ab initio} molecular dynamics (AIMD) simulations of the bulk $3 \times 3 \times 3$ supercell at room temperature, medium temperature (below melting point), high temperature (above melting point). In addition, snapshots were also obtained from $NVT$ AIMD simulations at room temperature at 90\% and 110\% of the equilibrium 0 K volume. Forty snapshots were extracted from each AIMD simulation at intervals of 0.1 $ps$; 
    \end{enumerate}
    \item Binary systems (Nb-Mo, Nb-Ta, Nb-W, Mo-Ta, Mo-W, Ta-W)
    \begin{enumerate}
        \item Solid solution structures constructed by partial substitution of $2 \times 2 \times 2$ bulk supercells of one element with the other element. Compositions of the form \ce{A_$x$B_$1-x$} were generated with $x$ ranging from 0 at\% to 100 at\% at intervals of 6.25 at\%.
    \end{enumerate}
    \item Ternary and quaternary systems (Nb-Mo-Ta, Nb-Mo-W, Mo-Ta-W, Nb-Ta-W, Nb-Mo-Ta-W)
    \begin{enumerate}
        \item Special quasi-random structures (SQS)\cite{zungerSpecialQuasirandomStructures1990} generated with ATAT code\cite{vandewalleAlloyTheoreticAutomated2002} using a $4 \times 4 \times 4$ bcc supercell. 
        \item Snapshots from $NVT$ AIMD simulations of the NbMoTaW SQS at 300, 1000, 3000 K.
    \end{enumerate}
\end{enumerate}

For the binary solid solution structures with each doping percentage, we performed a structure relaxation for all symmetrically distinct structures. Both the unrelaxed and relaxed structures were included in our data set. For the ternary and quaternary systems, our data set also includes structures generated by permuting the elements in the generated SQS, as well as the relaxed structures (including the intermediate structures) by optimizing the generated SQS. 
 
The test set of structures are generated by extracting additional snapshots from all previous AIMD simulations at intervals of 0.1 $ps$. We also generated additional binary solid solution structures by partial substitution of one element with the other element in a $2\times2\times1$ supercell. The substitution percentage ranges from 0 at$\%$ to 100 at$\%$ at intervals of 25 at$\%$. The total number of test structures is about 10$\%$ of that for training data.
 
\subsection{DFT calculations}
We performed the DFT calculations using the Perdew-Burke-Ernzerhof (PBE) \cite{perdewGeneralizedGradientApproximation1996} exchange-correlation functional and projector-augmented plane wave (PAW)\cite{blochlProjectorAugmentedwaveMethod1994} potentials as implemented in the Vienne \textit{ab initio} simulation package (VASP)\cite{kresseEfficientIterativeSchemes1996}. The plane-wave cutoff energy was 520 eV, and the $k$-point density was $4\times4\times4$ for $3\times3\times3$ supercells. The energy threshold for self-consistency and the force threshold for structure relaxation were 10$^{-5}$ eV and 0.02 eV/\AA, respectively. A single $\Gamma$  $k$ point was applied for non-spin-polarized AIMD simulations. However, we used the same parameters as the rest of the data for the energy and force calculations on the snapshots. The Python Materials Genomics (pymatgen)\cite{ongPythonMaterialsGenomics2013} library was used for all structure manipulations and analysis of DFT computations. The Fireworks software\cite{jainFireWorksDynamicWorkflow2015} was applied for the automation of calculations.

\subsection{SNAP model fitting}
 
The fitting workflow for the quaternary SNAP model is illustrated in Fig.\ref{fig:workflow}, in which we adopt the potential fitting workflow for elemental SNAP model developed in Ref.\citenum{chenAccurateForceField2017} as an optimization unit. This optimization unit contains two optimization loops. The inner loop optimizes the ML model parameters ($\beta$ in Fig.\ref{fig:workflow}) by mapping the descriptors (bispectrum coefficients) to DFT calculated formation energies and forces. The formation energies are defined as, $E_{form} = E^{TOT} - \sum_{el = Nb,Mo,Ta,W} N_{el}E_{el}$, where $E^{TOT}$ is DFT calculated total energy of the system; $N_{el}$ is the number of atoms in the system for the each type of element; $E_{el}$ is the energy per atom in the corresponding elemental bulk system. The hyperparameters are optimized in the outer loop by minimizing the difference between the model predicted material properties, i.e., elastic tensors, and the DFT computed values. These hyperparameters include the data weight ($\omega$ in Fig.\ref{fig:workflow}) from different data groups, and the parameters ($\alpha$ in Fig.\ref{fig:workflow}) used in bipectrum calculations, i.e., the radius cutoff $R_c$, and atomic weight $S_{atom}^k$. The fitting algorithm for each loop is the same as previous works\cite{chenAccurateForceField2017, liQuantumaccurateSpectralNeighbor2018} with the least-squares algorithm for inner loop and the differential evolution algorithm\cite{Optimize} for outer loop.
 
For the quaternary NbMoTaW alloy system, there are eight hyperparameters ($R_c^{Nb}$, $R_c^{Mo}$, $R_c^{Ta}$ ,$R_c^{W}$, $w_{atom}^{Nb}$, $w_{atom}^{Mo}$, $w_{atom}^{Ta}$, $w_{atom}^{W}$) in the bispectrum calculations, two for each element. A more-efficient step-wise optimization was performed. In the first step, we performed a series of independent optimization of the radius cutoff $R_c$ for each elemental SNAP model, i.e., Nb, Mo, Ta and W. The optimized radius cutoffs are then used as the radius cutoff for the quaternary NbMoTaW SNAP model. In the second step, a grid search was performed for the atomic weight for each element. We initially fix the data weight according to the number of data points for each data groups, and perform a quick grid search for the atomic weight of each element by only running the inner loop. The grid range is confined between 0 and 1.0 with interval 0.1 for each atomic weight. We then select the first ten combinations of the atomic weights of the four elements (see Table S2) with the best accuracy in energy and force predictions to conduct a full optimization, including the optimization of the data weight in the outer loop. The best model (with the highest accuracy in energies and forces) was selected from the ten fully optimized models.

\subsection{Atomistic simulations}

Atomistic simulations using the MPEA SNAP model were performed using the LAMMPS code.\cite{plimptonFastParallelAlgorithms1995} Specifically,
\begin{itemize}
    \item \textbf{Melting points.} The solid-liquid coexistence approach\cite{morrisMeltingLineAluminum1994} was used for melting temperature calculations. We use a $30\times 15\times 15$ bcc (13,500 atoms) supercell for each system to perform MD simulations under zero pressure at various temperatures. 1 fs was set for the time step. We performed the MD simulations for at least 100 ps to ensure the correct conclusion. The temperature, at which the initial solid and liquid phases were at equilibrium without interface motion, was identified as the melting point. 
    \item \textbf{Generalized stacking fault (GSF) energies.} The GSF energies were performed using a large supercell containing about 36,000 atoms. The supercell was set to be periodic along $<$111$>$ and $<$112$>$ directions in the $\{110\}$ plane and non-periodic along $<$110$>$ direction.
    \item \textbf{Dislocation core structure.} To study dislocation core structure and dynamics, we inserted a 1/2[111] screw disclocation with line direction $z$=[111], glide direction $x$=[11-2], and glide plane normal $y$=[-110] into a cylinder supercell with a radius 10 nm. The length of the dislocation line of one periodicity along $<$111$>$ direction is 8$b$ ($b$ is the Burgers vector) in the supercell. The quaternary cylinder supercell is constructed from a 54-atom SQS. The dislocation was inserted by deforming the atomic positions according to the  linear elasticity theory. Rigid boundary conditions were used by creating a layer of atoms fixed in their unrelaxed position outside of the inner cylinder region with radius 9 nm of mobile atoms. This method with this configuration has been used to study dislocation properties in previous works for bcc metals\cite{fellingerForcematchedEmbeddedatomMethod2010,segallAccurateCalculationsPeierls2001}. Similarly, a 1/2[111] edge dislocation with the line direction along z=[11-2] could be introduced inside the cylinder supercell. Energy minimization was performed using periodic boundary conditions along the dislocation line direction ($z$ direction) and fixed boundary conditions along the other two directions ($x$ and $y$ directions). To measure the critical resolved shear stress or Peierls stress for dislocation motion at T = 0 K, we applied increasing homogeneous shear strain in small increments and determined the stress value at which the dislocation moves from its initial position as the critical resolved shear stress. For MPEA, we record the largest shear stress within one periodicity.
    \item \textbf{Polycrystal simulations.} The initial polycrystal model was generated using the Voronoi tessellation method\cite{brostowConstructionVoronoiPolyhedra1978} implemented in the Atomsk code\cite{hirelAtomskToolManipulating2015}. We constructed a $11\times 11\times 11$ nm supercell and randomly inserted six GBs with average grain diameter about 7.5 nm. Periodic boundary conditions were imposed in all directions. At the GBs, pairs of atoms with distance smaller than 1.5 $\text{\AA}$ were removed. A hybrid MC/MD simulation was performed on the MPEA polycrystal for 1.5 ns. The MD simulations were carried out in NPT ensembles at 300 K and zero pressure. The time step was set to 5 fs. In the MC run, the sample was initially generated by randomly inserting the four elements to the supercell. Trial swaps were performed by randomly selecting an atom and replacing it with another species. The relative mole fractions of different elements are kept constant. The trial moves are accepted or rejected based on the Metropolis algorithm. A swap attempt was conducted at every MD step.
    \item \textbf{Stress-strain simulations.} Uniaxial compressive deformation simulations were carried out with a strain rate $5\times10^8$ $s^{-1}$. The time step was set to 1 fs, and simulations were performed under NPT ensemble at room temperature. 
\end{itemize}

\subsection{Chemical short-range-order parameters}
The definition of the pairwise multicomponent short-range order parameter is
\begin{equation}
    \alpha^k_{ij}=(p^k_{ij}-c_j)/(\delta_{ij}-c_j),
\end{equation}
where $k$ denotes the $k^{th}$ nearest-neighbor shell of the central atom $i$, $p^k_{ij}$ is the average probability of finding a $j$-type atom around an $i$-type atom in the $k^{th}$ shell, $c_j$ is the average concentration of $j$-type atom in the system, and $\delta_{ij}$ is the Kronecker delta function. For pairs of the same species (i.e., $i = j$), a positive $\alpha^k_{ij}$ means the tendency of attraction in the $k^{th}$ shell and a negative value suggests the tendency of repulsion. For pairs of different elements (i.e., $i \neq j$), it is the opposite. A negative $\alpha^k_{ij}$ suggests the tendency of $j$-type clustering in the $k^{th}$ shell of an $i$-type atom, while a positive value means the repulsion.

\section{Data availability}

To ensure the reproducibility and use of the models developed in this paper, all data (structures, energies, forces, etc.) used in model development as well as the final fitted model coefficients have been published in an open repository (https://github.com/materialsvirtuallab/snap). 

\section{Code availability}

The DFT calculations were performed with the Vienna ab initio simulation package. The LAMMPS package was used to perform MD/MC simulations. All the other codes that support the findings of this study are available from Dr. Xiang-Guo Li (email: xil110@ucsd.edu) upon reasonable request.

\section{Acknowledgement}

This work is funded by the Office of Naval Research under Grant number N00014-17-S-BA13. We thank Dr. Shuozhi Xu for helpful discussions. The authors also acknowledge computational resources provided by the Triton Shared Computing Cluster (TSCC) at the University of California, San Diego, and the Extreme Science and Engineering Discovery Environment (XSEDE) supported by National Science Foundation under grant no. ACI-1053575.

\section{AUTHOR CONTRIBUTIONS}
X.-G. L. performed potential model training, performance evaluation and alloy mechanical property investigations.
C. C., H. Z. and Y. Z. helped with the analyses in alloy mechanical property investigations.
S.P. O. is the primary investigator and supervised the entire project.
All authors contributed to the writing and editing of the paper.

\section{COMPETING INTERESTS}
The authors declare no Competing Financial or Non-Financial Interests.

\section{ADDITIONAL INFORMATION}
\textbf{Supplementary information} accompanies the paper on the npj Computational
Materials website .

\end{document}